\documentclass[pra,aps,twocolumn,superscriptaddress,showpacs,nofootinbib]{revtex4}

\usepackage{amsmath,amssymb,amsbsy,graphicx,bm,subfigure}

\newcommand{\eq}[1]{Eq.~(\ref{#1})}

\begin{document}

\title{Nonclassical correlations in continuous-variable non-Gaussian Werner states}


\date{\today}

\begin{abstract}
We study nonclassical correlations beyond entanglement in a family of two-mode non-Gaussian states which represent the continuous-variable counterpart of two-qubit Werner states. We evaluate quantum discord and other quantumness measures obtaining  exact analytical results in special instances, and upper and lower bounds in the general case. Non-Gaussian measurements such as photon counting are in general necessary to solve the optimization in the definition of quantum discord, whereas Gaussian measurements are strictly suboptimal for the considered states. The gap between Gaussian and optimal non-Gaussian conditional entropy is found to be proportional to a measure of non-Gaussianity in the regime of low squeezing, for a subclass of continuous-variable Werner states. We further study an example of a non-Gaussian state which is positive under partial transposition, and whose nonclassical correlations stay finite and small even for infinite squeezing. Our results pave the way to a systematic exploration of the interplay between nonclassicality and non-Gaussianity in continuous-variable systems, in order to gain a deeper understanding of ---and to draw a bigger advantage from--- these two important resources for quantum technology.
\end{abstract}

\author{Richard Tatham}
\affiliation{School of Physics and Astronomy, University of St.~Andrews, North Haugh, St.~Andrews, Fife, KY16 9SS, Scotland}
\author{Ladislav Mi\v{s}ta, Jr.}
\affiliation{Department of Optics, Palack\' y University, 17.~listopadu 12,  771~46 Olomouc, Czech Republic}
\author{Gerardo Adesso}
\affiliation{School of Mathematical Sciences, University of Nottingham, University Park, Nottingham NG7 2RD, United Kingdom}
\author{Natalia Korolkova}
\affiliation{School of Physics and Astronomy, University of St.~Andrews, North Haugh, St.~Andrews, Fife, KY16 9SS, Scotland}

\maketitle

\section{Introduction}

Two decades after pioneering contributions of the likes of quantum cryptography \cite{ekert} and teleportation \cite{telep}, quantum information science has nowadays acquired a status of maturity \cite{nielsen}, having witnessed theoretical and experimental advances which demonstrated in several ways the power of quantum technology \cite{techNS}. Nonetheless, fundamental questions of inherent practical relevance remain open, broadly concerning the proper identification of the ultimate {\it resources} behind such a power. Paradigmatic instances of systems suitable for quantum protocols have been, for instance, registers of pure qubits for discrete-variable (``digital'') quantum computation and cryptography \cite{nielsen,naturebennett}, and multimode Gaussian states of radiation fields for continuous-variable (``analog'') quantum communication and information processing \cite{cvreview}. Entanglement \cite{entanglement} has been the crucial resource on which most studies have focused so far as its presence is crucial for better-than-classical communication performances. Recently, however, there has been more of a push for research in quantum information to go beyond the boundaries of its first generation. In particular, the need to consider more realistic setups where mixedness affects quantum computations in the discrete variable framework \cite{dqc1}, and the need to go beyond the nutshell of Gaussian states and operations to achieve universality in continuous variable computation \cite{menicucci} have both been strong motivations. Nonclassicality (i.e. the quantumness of correlations beyond and without entanglement) \cite{merali} and non-Gaussianity \cite{nongauss} are currently under the limelight due to being recognized as ``power-ups'' for quantum technology, in particular in applications such as quantum computation \cite{dqc1,menicucci}, quantum communication \cite{discordcommun,nongausscommun} and metrology \cite{discordmetro,nongaussmetro}.

In this paper we investigate the nonclassicality of correlations in continuous-variable (CV) systems beyond Gaussian states.  Nonclassical correlations, including and beyond entanglement, can be quantified for instance by the {\em quantum discord} \cite{Olivier_01,Vedral_01}, a measure that aims at capturing more general signatures of quantumness in composite systems. They are associated e.g.~with the non-commutativity of quantum observables and with the fact that local measurements generally induce some disturbance on quantum states, apart from very special cases in which those states admit a fully classical description. The interest in quantum discord has risen significantly since it was recognized as the potential resource behind the quantum speed-up in certain mixed-state models of quantum computing such as the DQC1 \cite{dqc1}, in which entanglement is negligible or strictly vanishing \cite{dattaetalqc}. Quantum discord \cite{Olivier_01}, along with other similar nonclassicality measures \cite{cavesrossignoli} such as the (ameliorated) measurement-induced disturbance \cite{Luo_08a,Wu_09,Girolami_10} and distance-based quantifiers such as the geometric discord \cite{dakic} and the relative entropy of quantumness \cite{req,Piani_11}, typically feature nontrivial optimizations in their definitions. In particular, it is necessary to identify the least disturbing measurement to be applied on one or more subsystems to extract those nonclassical correlations, rendering their exact computation a formidable task. Closed formulae are available for the quantum discord of a special subclass of two-qubit states \cite{discordqubits} and of general two-mode Gaussian states under the restriction of Gaussian measurements \cite{gdiscord}.

We consider here a family of two-mode states that are the CV counterparts of two-qubit Werner states \cite{werner}. They are non-Gaussian states obtained as mixtures of two Gaussian states \cite{Mista_02}, an entangled two-mode squeezed state and a two-mode thermal product state, and find applications in CV quantum cryptography \cite{Lund_06}. Studying their nonclassicality beyond entanglement is particularly interesting from a fundamental point of view, as they offer a unique testbed to compare the role of Gaussian versus non-Gaussian measurements to extract correlations with minimum disturbance from general two-mode CV states. Gaussian states and operations are known to satisfy sharp extremality properties \cite{extra} in the space of all CV states. However, such results do not apply to quantum discord, rendering the understanding of the structure of (non)classical correlations even subtler in CV systems compared to finite-dimensional systems, for which general results are known instead \cite{ferraro}. We show that the states analyzed here represent instances of quantum states carrying {\it genuine} CV nonclassical non-Gaussian correlations, for which the optimization in the discord is achieved {\it only} by an infinite-dimensional component, hence genuinely CV, non-Gaussian measurement.\footnote{One might argue, otherwise, that e.g.~any two-qubit state is an example of a non-Gaussian state; for them, however, the computation of discord involves two-component measurements, which are certainly non-Gaussian but span just a two-dimensional space. Here we focus instead on CV states and measurements spanning the whole infinite-dimensional Hilbert spaces, and not just finite-dimensional truncations of it.}
In the following we prove that for a particular type of the non-Gaussian CV Werner state \cite{Mista_02} (obtained by mixing a two-mode squeezed state with the vacuum)
quantum discord can be computed {\it analytically}. The global optimization in its definition is achieved exactly by photon counting, and such a measurement is strictly less disturbing than any Gaussian measurement. The gap between optimal Gaussian measurements (homodyne detection) and optimal non-Gaussian measurements (photon counting) is quantified by the difference of the associated conditional entropies. The gap is then compared with an entropic measure for the non-Gaussianity of the states \cite{Genoni_08}, and found to be proportional to it in the low squeezing regime. The exactly computed discord turns out to be equal to other measures for the quantumness of correlations as well in this special case \cite{Girolami_10,Piani_11}.   Next, we analyze discord in the most general CV Werner state, ideally complementing the analysis of the two-qubit Werner state performed originally in \cite{Olivier_01} (see also \cite{discordqubits}). We show that photon counting provides in general a (not necessarily optimal) upper bound on quantum discord, that coincides with the measurement-induced disturbance \cite{Luo_08a}, and we derive a  lower bound on discord as well. Finally, we give an example of a CV state which is positive under partial transposition (PPT \cite{PPT}), and show that it carries ``weak'' nonclassical correlations, signaled by analytically computable upper and lower bounds on discord which are close to each other and stay small and finite even for infinite squeezing. Our results provide insights into the relationship between the general quantumness of correlations, entanglement distillability and separability in CV systems outside the Gaussian scenario. From a practical perspective, our results identify the key role of non-Gaussian measurements such as photon counting to access and extract all nonclassical correlations in general CV states, even in the particular case of non-Gaussian states with a positive-everywhere Wigner function such as the mixtures of Gaussian states studied here.

The paper is organized as follows.
In Sec.~\ref{secbasic} we set up notation and recall the definitions of quantum discord \cite{Olivier_01} and of the CV Werner states \cite{Mista_02}. In Sec.~\ref{secexact} we study a special type of CV Werner state and calculate its discord (and other nonclassicality indicators) exactly, proving that photon counting constitutes a globally optimal measurement strategy, and issuing comparisons between the degree of non-Gaussianity \cite{Genoni_08} and the amount by which Gaussian measurements fail to be optimal on the considered class of states. In Sec.~\ref{secgeneral} we address the general case and provide exact lower and upper bounds for the quantum discord of arbitrary CV Werner states. In Sec.~\ref{secppt} we study a special type of CV state obtained by partial transposition from a CV Werner state, and investigate its degree of nonclassical correlations in different entanglement regimes. We draw our conclusions in Sec.~\ref{secfinal}.

\section{Basic notation and definitions}\label{secbasic}
\subsection{Quantum discord}\label{secdiscord}
Quantum discord is a measure of the nonclassicality of correlations in a bipartite quantum state $\rho_{AB}$
defined as a difference \cite{Olivier_01}
\begin{equation}\label{D1}
\mathcal{D}(\rho_{AB})={\cal I}_{q}(\rho_{AB})-\mathcal{J}(\rho_{AB})
\end{equation}
between the quantum generalizations of two classically equivalent
expressions for the mutual information, namely the quantum mutual information
\begin{equation}\label{I} {\cal I}_{q}(\rho_{AB})={\cal
S}(\rho_{A})+{\cal S}(\rho_{B})-{\cal S}(\rho_{AB}),
\end{equation}
where ${\cal S}(\rho)=-\mbox{Tr}(\rho\ln\rho)$ is the von Neumann
entropy, and the so called one-way classical correlation \cite{Vedral_01}
\begin{equation}\label{J}
\begin{split}
{\cal J}(\rho_{AB})={\cal S}(\rho_{A}) - \inf_{\{\Pi_i\}} {\cal
H}_{\{\Pi_i\}}(A|B),
\end{split}
\end{equation}
with ${\cal H}_{\{\Pi_i\}}(A|B){\equiv}\sum_{i}p_{B}(i){\cal
S}(\rho_{A|i})$ being the conditional entropy of $A$ given a positive operator valued measurement (POVM)
$\{\Pi_{B}(i)\}$ has been performed on $B$. 
Here $\rho_{A,B}=\mbox{Tr}_{B,A}[\rho_{AB}]$ are reduced states of
subsystems $A$ and $B$, respectively, $\rho_{A|i} =
\mbox{Tr}_B[\Pi_{B}(i)\rho_{AB}]/p_{B}(i)$ is the conditional state obtained
upon detecting the POVM element $\Pi_{B}(i)$ on $B$ and
$p_{B}(i)=\mbox{Tr}[\Pi_{B}(i)\rho_{AB}]$ is the corresponding
probability. Without loss of generality
one can restrict to rank 1 POVMs in what follows \cite{Datta_2011}. 

Subtracting \eq{J} from \eq{I} one gets for quantum
discord the following formula:
\begin{equation}\label{D2}
\mathcal{D}(\rho_{AB})={\cal S}(\rho_{B})-{\cal
S}(\rho_{AB})+\inf_{\{\Pi_i\}}{\cal
H}_{\{\Pi_i\}}(A|B).
\end{equation}
Note that quantum discord is generally asymmetric under the change
$A\leftrightarrow B$, but in this paper we will consider only symmetric states
for which the discord is naturally invariant under the choice of the measured subsystem.

The hard step in the evaluation of the quantum discord
(\ref{D2}) is the optimization of the conditional entropy ${\cal
H}_{\{\Pi_i\}}(A|B)$ over all POVMs. Although this cannot be done in
full generality analytically even for two qubits \cite{discordqubits}
the problem is tractable using analytical tools for certain subsets
of states and POVMs, e.g., for Gaussian ones \cite{gdiscord}.
However, the question as to whether Gaussian measurements are globally optimal for the extraction of nonclassical correlations from general CV quantum states is essentially open and has not been settled even for Gaussian states themselves.\footnote{In this respect it is known that, for certain Gaussian states, non-Gaussian measurements are needed to minimize two-way quantifiers of nonclassical correlations such as the ameliorated measurement-induced disturbance \cite{gamid}. For the quantum discord itself, Gaussian POVMS are instead conjectured to be optimal among all possible CV measurements \cite{gdiscord,Datta_10}.}

\subsection{CV Werner state}\label{secCVW}
We consider the CV Werner state \cite{Mista_02} defined as
\begin{equation}\label{rho}
\rho=p\,|\psi(\lambda)\rangle\langle \psi(\lambda)|+(1-p)\,{\rho}^{\rm th}_{A}(\mu)\otimes{\rho}^{\rm th}_{B}(\mu),
\end{equation}
where $0\leq p\leq 1$,
\begin{equation}\label{psinopa}
|\psi(\lambda)\rangle=\sqrt{1-\lambda^2}\sum_{n=0}^{\infty}
\lambda^{n}|n,n\rangle_{AB}.
\end{equation}
is the two-mode squeezed vacuum state with $\lambda=\tanh r$
($r$ is the squeezing parameter) and
\begin{equation}\label{rhoth}
{\rho}^{\rm th}_{j}(\mu)=(1-\mu^{2})\sum_{n=0}^{\infty}
\mu^{2n}|n\rangle_{j}\langle n|,\quad j=A,B
\end{equation}
is the thermal state with $\mu^{2}=\langle n_{j}\rangle/(1+\langle n_{j}\rangle)$, where
$\langle n_{j}\rangle$ is the mean number of
thermal photons in mode $j$.  We observe that the infinite-dimensional state (\ref{rho}) also
possesses the same structure as $d$-dimensional states invariant under the maximal
commutative subgroup of $U(d)$ introduced in \cite{Chruscinski_06}.

\section{Special case ${\bm{\mu=0}}$}\label{secexact}

First, let us consider the simplest special case of a Werner
state with $\mu=0$ which gives using \eq{rho}
\begin{equation}\label{rho0}
\rho_{0}=p\,|\psi(\lambda)\rangle\langle \psi(\lambda)|+(1-p)\,|00\rangle\langle00|\,,
\end{equation}
representing just a mixture of a two-mode squeezed vacuum state with the vacuum.
For $p>0$ the partially transposed matrix $\rho_{0}^{T_{A}}$ (obtained by transposing $\rho_0$ with respect to the degrees of freedom of subsystem $A$ only) has negative eigenvalues \cite{Mista_02} and therefore, according to the PPT criterion \cite{PPT}, the state (\ref{rho0}) is entangled.
Note that the state (\ref{rho0}) has been further studied in \cite{Lund_06,carles} from the point of view of its entanglement properties, as measured by the negativity \cite{nega}, highlighting its applications for quantum key distribution.

\subsection{Exact calculation of quantum discord}
In order to calculate the entropies arising in the expression of quantum discord
(\ref{D2}) we need to determine the eigenvalues of the reduced state $\rho_{0,B}$,
the global state (\ref{rho0}) and the conditional state $\rho_{A|i} =
\mbox{Tr}_B[\Pi_{B}(i)\rho_{0}]/p_{B}(i)$. The latter two states attain the form
\begin{eqnarray}\label{sigma}
\sigma=\zeta_1\,|\phi_{1}\rangle\langle \phi_{1}|+\zeta_2\,|\phi_{2}\rangle\langle\phi_{2}|,
\end{eqnarray}
where $\zeta_1+\zeta_2=1$ and $|\phi_{1,2}\rangle$ are generally nonorthogonal
normalized pure state vectors. The state (\ref{sigma}) has at most
two-dimensional support spanned by vectors $|\phi_{1,2}\rangle$
corresponding to eigenvalues $\nu_{1,2}$ that read as
\begin{eqnarray}\label{nu}
\nu_{1,2}=\frac{1\pm\sqrt{1-4\zeta_1\zeta_2\left(1-|\langle\phi_{1}|\phi_{2}\rangle|^2\right)}}{2}.
\end{eqnarray}
On inserting the eigenvalues (\ref{nu}) into the formula for the
von Neumann entropy
\begin{equation}\label{S}
\mathcal{S}(\sigma)=-\sum_{i=1}^{2}\nu_{i}\ln\nu_{i}
\end{equation}
we get analytically the entropy of the state (\ref{sigma}).

Returning back to the state (\ref{rho0}) we get, in particular,
$|\phi_{1}\rangle=|\psi(\lambda)\rangle,|\phi_{2}\rangle=|00\rangle,\zeta_1=p$
and $\zeta_2=1-p$;  the eigenvalues thus amount to
\begin{eqnarray}\label{nurho0}
\nu_{1,2}=\frac{1\pm\sqrt{1-4p(1-p)\lambda^2}}{2}.
\end{eqnarray}
Hence, we can immediately calculate the entropy ${\cal
S}(\rho_{0})$ using formula (\ref{S}),
\begin{eqnarray}\label{srho0}
{\cal S}(\rho_0) &=& \mbox{$-\left(\frac{1 + \sqrt{1-4p(1-p)\lambda^2}}{2}\right) \ln
\left(\frac{1 + \sqrt{1-4p(1-p)\lambda^2}}{2}\right)$} \nonumber \\
& & \mbox{$-\left(\frac{1 - \sqrt{1-4p(1-p)\lambda^2}}{2}\right) \ln
\left(\frac{1 - \sqrt{1-4p(1-p)\lambda^2}}{2}\right).$} \nonumber \\
\end{eqnarray}
Tracing, further, the state
(\ref{rho0}) over mode $A$ yields a diagonal reduced state
\begin{equation}\label{rho0B}
\rho_{0,B}=p{\rho}^{\rm
th}_{B}(\lambda)+(1-p)|0\rangle_{B}\langle 0|,
\end{equation}
with ${\rho}^{\rm th}_{B}$ given in Eq. (\ref{rhoth}), possessing the eigenvalues

\begin{eqnarray}\label{nurho0B}
\tilde{\nu}_{0}=1-p\lambda^2,\quad
\tilde{\nu}_{n>0}=p(1-\lambda^2)\lambda^{2n},
\end{eqnarray}
for $n \in {\mathbb{N}}$. Making use of the definition (\ref{S}) we get also the entropy of
the reduced state in the form:
\begin{eqnarray}\label{SrhoB0}
\mathcal{S}(\rho_{0,B})&=&-\left\{\ln(1-p\lambda^2)+p\lambda^2\ln\left[\frac{p(1-\lambda^2)}{1-p\lambda^2}\right]\right.\nonumber\\
&&\left.+\frac{2p\lambda^2\ln\lambda}{1-\lambda^2}\right\}.
\end{eqnarray}

\begin{figure}[tb]
\includegraphics[width=8.5cm]{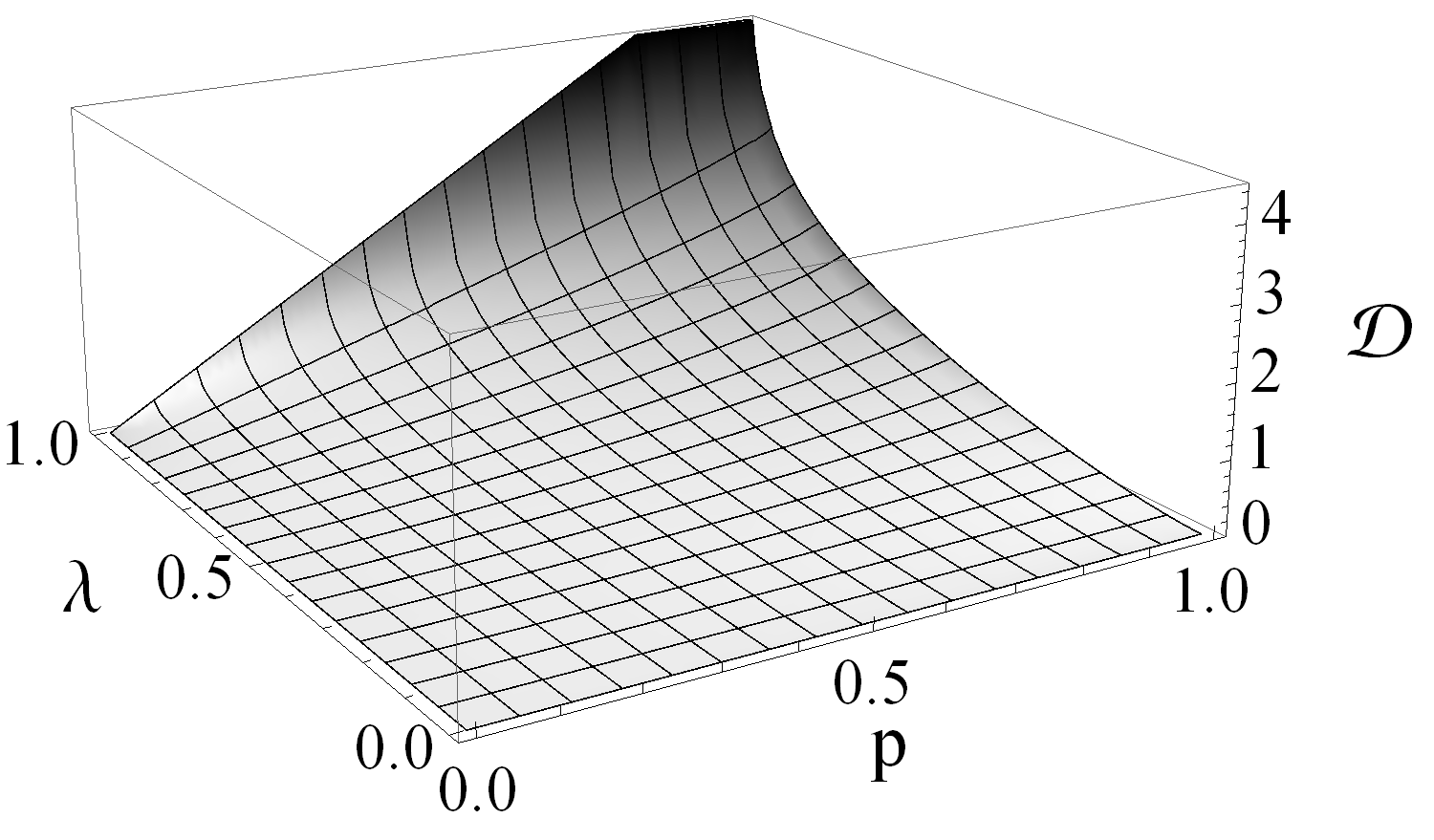}
\caption{Quantum discord ${\cal D}$ [Eq.~(\ref{Drho0})] versus the probability $p$ and the squeezing factor $\lambda$ for the CV Werner state $\rho_0$ [Eq.~(\ref{rho0})]. All the quantities plotted are dimensionless.}
\label{fig13d}
\end{figure}

We are left with the minimization
of the conditional entropy ${\cal H}_{\{\Pi_i\}}(A|B)$ over all POVMs.
The state (\ref{rho0}) is constructed in such a way that we can ``guess'' the optimal measurement
before attempting to solve this anyway hopeless task with brute-force. A close look at the
state reveals that the optimal measurement on mode $B$ is just photon counting (i.e., in this case, projection onto the eigenstates of $\rho_{0,B}$) characterized by the
set of projectors $\{\Pi_{B}\left(m\right)=|m\rangle_{B}\langle m|\}$, where $|m\rangle$ is the $m$-th Fock state.
Clearly, if we detect $m$ photons in mode $B$ in the state (\ref{rho0}), the conditional state
$\rho_{A|m}$ is simply a {\it pure} Fock state $|m\rangle_{A}\langle m|$ which implies, immediately,
that the conditional entropy attains its minimal possible value ${\cal H}_{\{\Pi(m)\}}(A|B)=0$. This proves rigorously that
photon counting is the globally optimal measurement strategy for the state $\rho_0$, thus giving the discord exactly equal to
\begin{equation}\label{Drho0}
\mathcal{D}(\rho_{0})={\cal S}(\rho_{0,B})-{\cal S}(\rho_{0}),
\end{equation}
where the involved entropies are defined in equations \eqref{srho0} and \eqref{SrhoB0}. The discord is an increasing function of both $\lambda$ and $p$ and is plotted in Fig.~\ref{fig13d}. Note, that the considered Werner state (8)
belongs to the class of maximally correlated states for which
Eq. \eqref{Drho0} can be proved alternatively \cite{Cornelio_11} using the duality
relation between classical correlations and entanglement of
formation \cite{Koashi_04}.

\subsection{Discord versus Gaussian discord and the non-Gaussianity of the state}

A relevant question is whether photon counting is the only measurement realizing the global minimum in the evaluation of quantum discord, or, more specifically, whether there can exist Gaussian POVMs that attain instead an equally optimal performance on the state \eqref{rho0}. The answer, as we argue in the following, is that nonclassical correlations in the state (\ref{rho0}) captured by discord (\ref{Drho0}) cannot be extracted equally well by any Gaussian measurement. In other words, Gaussian discord ($\equiv {\cal D}^{G}(\rho_{0})$), defined by Eq.~(\ref{D2}) by imposing the minimization be restricted only to Gaussian measurements \cite{gdiscord}, is just a strictly larger upper bound on the true discord (\ref{Drho0}) for the mixed non-Gaussian CV Werner state \eqref{rho0}.
To show this, consider the following  Gaussian POVM \cite{Fiurasek_07} consisting of elements
\begin{eqnarray}\label{GPOVM}
\Pi(\alpha)=\frac{1}{\pi}|\alpha,\xi\rangle\langle\alpha,\xi|,
\end{eqnarray}
where
\begin{equation}
|\alpha,\xi\rangle\equiv D(\alpha)S(\xi)|0\rangle,
\end{equation}
with $\xi=t {\rm e}^{i2\varphi}$, is a pure normalized momentum-squeezed vacuum state
with squeezing parameter $t\in\ [0,\infty)$, that is rotated counterclockwise
by a phase $\varphi\in [0,\pi)$ and that is subsequently displaced  by $\alpha\in\mathbb{C}$.
Here, $D(\alpha)=\mbox{exp}(\alpha a^{\dag}-\alpha^{\ast}a)$ is the displacement operator and
$S(\xi)=\mbox{exp}\left\{\left[\xi (a^{\dag})^2-\xi^{\ast}a^2\right]/2\right\}$
is the squeezing operator. Note that $t=0$ corresponds to heterodyne detection,
whereas homodyne detection is obtained in the limit
$t\rightarrow\infty$. If the POVM element $\Pi(\alpha)$ is detected on mode $B$ of the
two-mode squeezed vacuum state (\ref{psinopa}), then mode $A$ collapses into the 
state $|\beta,\omega=s{\rm e}^{-i2\varphi}\rangle$, which is a pure momentum-squeezed state
with squeezing parameter
\begin{equation}\label{s}
s=\frac{1}{2}\ln\left[\frac{1+{\rm e}^{2t}\cosh(2r)}{\cosh(2r)+{\rm e}^{2t}}\right]
\end{equation}
that is rotated clockwise by phase $\varphi$ and that is displaced by
\begin{equation}\label{beta}
\beta=\frac{\sinh(2r)}{2}\left[\left(z_{+}+z_{-}\right)\alpha^{\ast}+\left(z_{+}-z_{-}\right){\rm e}^{-i2\varphi}\alpha\right],
\end{equation}
where $z_{\pm}=\left[\cosh(2r)+\mbox{exp}\left(\pm 2t\right)\right]^{-1}$.
Thus, the obtained conditional state $\rho_{A|\alpha}=\mbox{Tr}_B\left[\Pi_{B}\left(\alpha\right)\rho_{0}\right]/q\left(\alpha\right)$
is again a convex mixture of the form (\ref{sigma}) where $|\phi_{1}\rangle=|\beta,\omega\rangle$, $|\phi_{2}\rangle=|0\rangle$ and
\begin{eqnarray}\label{AB2}
{\zeta_1}_{\alpha}=\frac{pu(\alpha)}{\pi q(\alpha)},\quad {\zeta_2}_{\alpha}=\frac{(1-p)v(\alpha)}{\pi q(\alpha)}.\nonumber\\
\end{eqnarray}
\begin{figure}[tb]
\includegraphics[width=7.5cm]{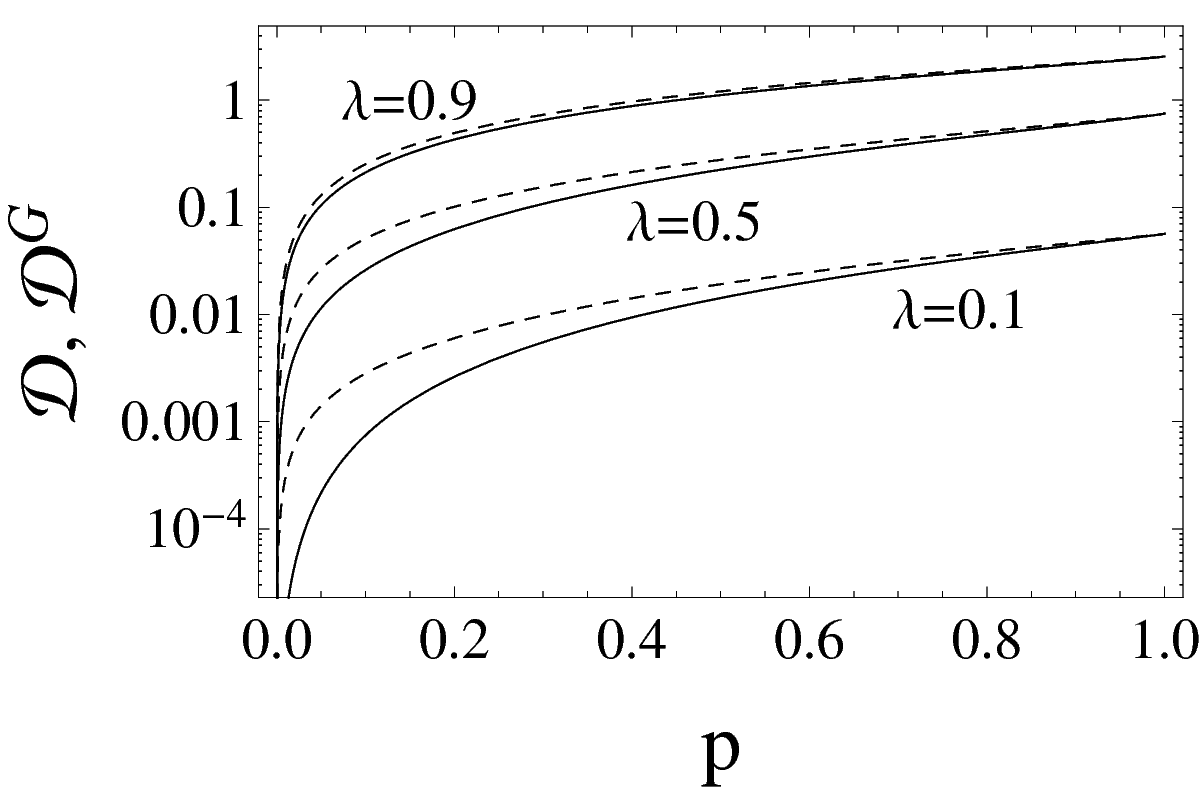}
\caption{Logarithmic plot of quantum discord (solid curve) and Gaussian quantum discord (dashed curve)
versus the probability $p$ for the CV Werner state (\ref{rho0}) with (from bottom to top) $\lambda=0.1, 0.5, 0.9$. All the quantities plotted are dimensionless.}\label{fig1}
\end{figure}
Here $q(\alpha)=\left[pu(\alpha)+(1-p)v(\alpha)\right]/\pi$ is the
probability density of obtaining the measurement outcome $\alpha$,
$u(\alpha)=\langle\alpha,\xi|\rho_{B}^{\rm
th}(\lambda)|\alpha,\xi\rangle$, where $\rho_{B}^{\rm
th}(\lambda)$ is given in Eq.~(\ref{rhoth}), and
$v(\alpha)=|\langle 0|\alpha,\xi\rangle|^2$. The latter function
$v(\alpha)$ can be computed straightforwardly using the formula
\cite{Perina_91}
\begin{equation}\label{overlap}
\langle0|\alpha,\xi=t{\rm e}^{i2\varphi}\rangle=\frac{{\rm e}^{-\frac{|\alpha|^2}{2}+\frac{\tanh
(t)}{2}{\rm e}^{i2\varphi}\alpha^{\ast2}}}{\sqrt{\cosh (t)}}.
\end{equation}
The overlap $|\langle\phi_{1}|\phi_{2}\rangle|^2=|\langle0|\beta,\omega=s{\rm e}^{-i2\varphi}\rangle|^2$ appearing in
eigenvalues (\ref{nu}) can be calculated exactly along the same lines.
Expressing finally the thermal state $\rho_{B}^{\rm th}(\lambda)$ as
\begin{equation}\label{rhothlambda}
\rho_{B}^{\rm th}(\lambda)=\frac{1-\lambda^{2}}{\pi\lambda^{2}}\int_{\mathbb{C}}{\rm e}^{-\frac{1-\lambda^2}{\lambda^2}|\varsigma|^2}|\varsigma\rangle\langle\varsigma|d^{2}\varsigma
\end{equation}
we get the function $u(\alpha)$ in the form
\begin{equation}\label{upom}
u(\alpha)=\frac{1-\lambda^{2}}{\pi\lambda^{2}}\int_{\mathbb{C}}{\rm e}^{-\frac{1-\lambda^2}{\lambda^2}|\varsigma|^2}
|\langle0|\alpha-\varsigma,\xi\rangle|^{2}d^{2}\varsigma,
\end{equation}
where we used the property of displacement operators $D(-\varsigma)D(\alpha)=\mbox{exp}[(\varsigma^{\ast}\alpha-\varsigma\alpha^{\ast})/2]D(\alpha-\varsigma)$.
Using once again the formula (\ref{overlap}) to express the overlap $|\langle0|\alpha-\varsigma,\xi\rangle|^{2}$ and performing the integration over $\varsigma$ we arrive at the formula
\begin{eqnarray}\label{u}
u(\alpha)&=&\frac{1-\lambda^{2}}{\cosh(t)\sqrt{1-\lambda^{4}\tanh^{2}(t)}} \\
&\times& \exp\Bigg\{-\frac{\left(1-\lambda^2\right)\left[1-\lambda^2\tanh^{2}(t)\right]}
{1-\lambda^{4}\tanh^{2}(t)}|\alpha|^{2}\nonumber \\
& & +\frac{\left(1-\lambda^2\right)^{2}\tanh(t)}{2\left[1-\lambda^{4}\tanh^{2}(t)\right]}
\left({\rm e}^{-i2\varphi}\alpha^2+{\rm e}^{i2\varphi}\alpha^{\ast 2}\right)\Bigg\}. \nonumber
\end{eqnarray}
Substituting the obtained explicit expressions for functions $u(\alpha), v(\alpha)$ and $q(\alpha)$
into Eqs.~(\ref{AB2}) and using the explicit expression for the overlap
$|\langle\phi_{1}|\phi_{2}\rangle|^2=|\langle0|\beta,\omega=s{\rm e}^{-i2\varphi}\rangle|^2$ we get
from Eq.~(\ref{nu}) the eigenvalues and hence the entropy
${\cal S}\left(\rho_{A|\alpha}\right)$ of the conditional state
$\rho_{A|\alpha}$. Subsequent averaging of the entropy over the density $q(\alpha)$
finally yields the Gaussian conditional entropy
\begin{equation}\label{Halpha}
{\cal
H}_{\{\Pi(\alpha)\}}^{G}(A|B)=\int_{\mathbb{C}}q(\alpha){\cal S}\left(\rho_{A|\alpha}\right)
d^{2}\alpha
\end{equation}
as a function of the squeezing parameter $t$ and phase $\varphi$ of
the Gaussian measurement (\ref{GPOVM}). Due to the complicated dependence of the conditional
entropy ${\cal S}\left(\rho_{A|\alpha}\right)$ on $\alpha$, the remaining integration over
the complex plane $\mathbb{C}$, where $d^{2}\alpha\equiv d(\mbox{Re}\alpha)d(\mbox{Im}\alpha)$,
has to be performed numerically. Likewise, minimization of the entropy (\ref{Halpha})
with respect to variables $t$ and $\varphi$ also requires numerics. This analysis reveals that, within the Gaussian POVM set,
the entropy is minimized by homodyne detection on mode $B$. The resulting plots of Gaussian discord and the true quantum discord (\ref{Drho0}) (the latter obtained by photon counting on $B$) are shown in Fig.~\ref{fig1}. The figure clearly shows that apart from trivial cases $p=0,1$ the Gaussian discord
is always strictly larger than the discord (\ref{Drho0}), meaning that general Gaussian measurements are strictly suboptimal (or, in other words, non-minimally disturbing) for the extraction of nonclassical correlations in the non-Gaussian state \eqref{rho0}.

\begin{figure*}[tb]
\centering
\subfigure[]{\includegraphics[width=7.5cm]{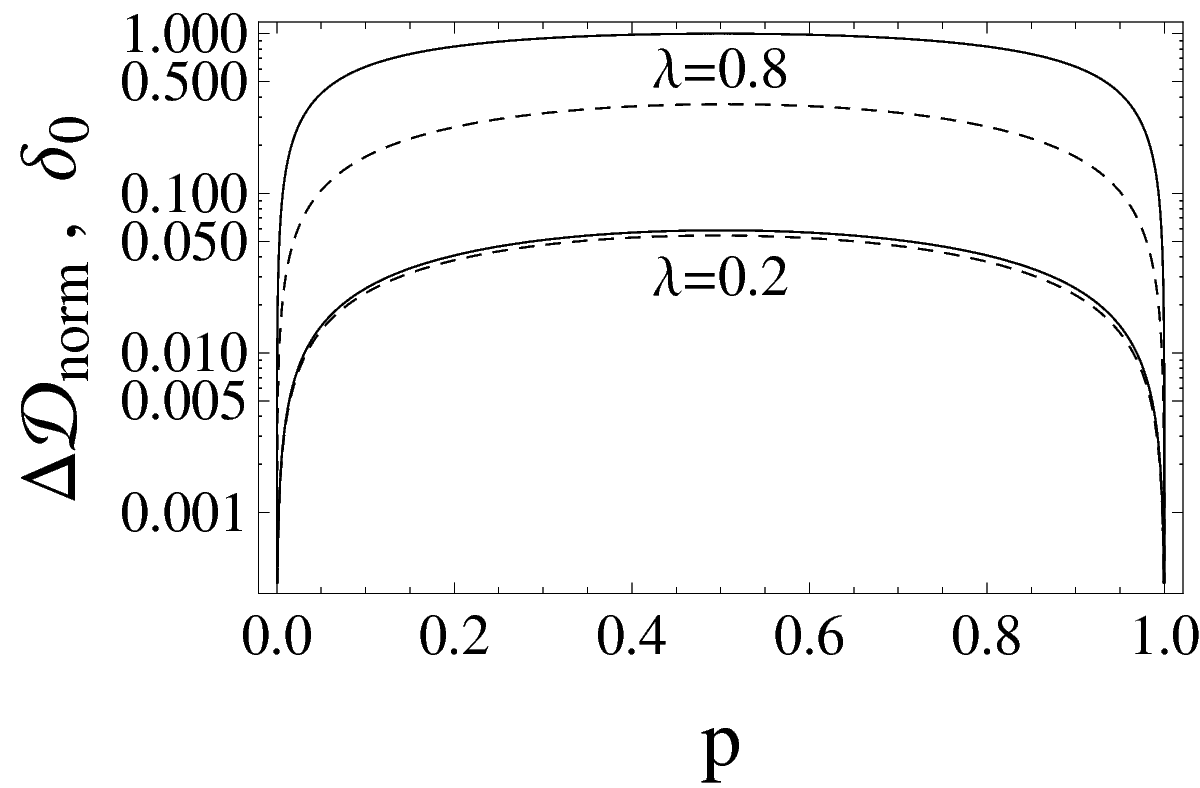}\label{figapp}}\hspace*{1.5cm}
\subfigure[]{\includegraphics[width=7.5cm]{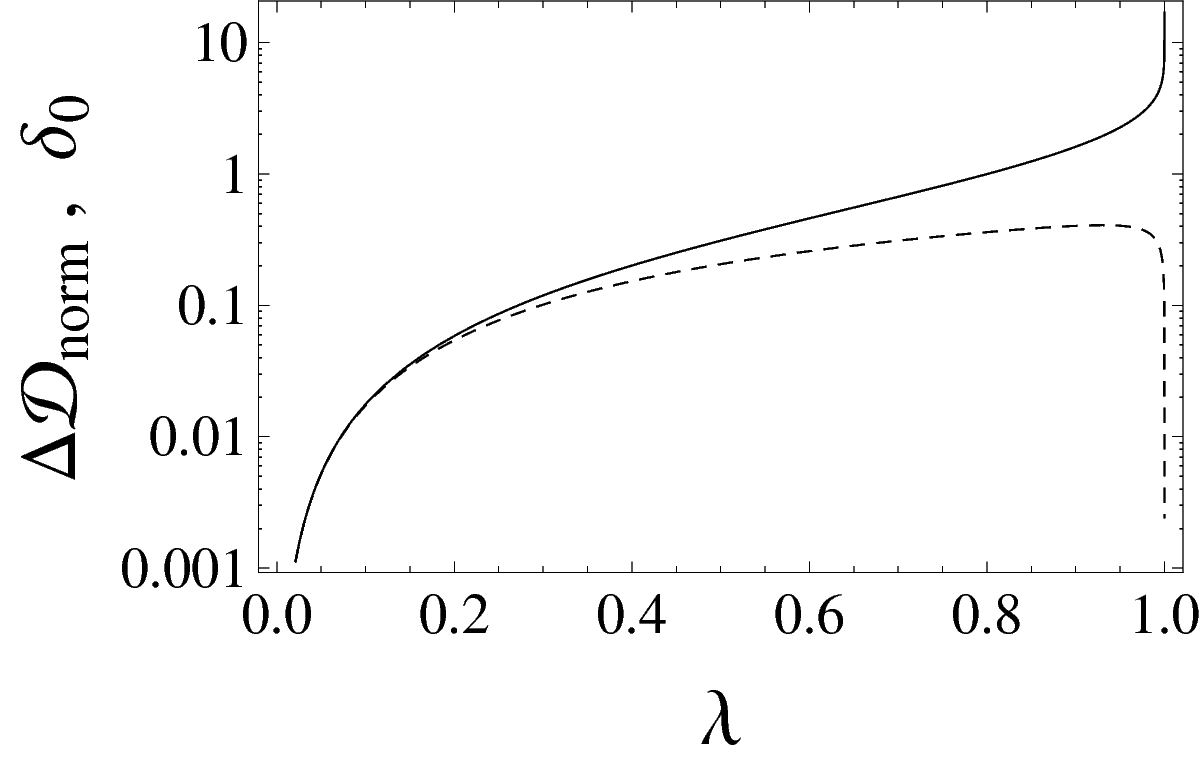}\label{figapl}}
\caption{Normalized gap $\Delta{\cal D}_{\text{norm}}$ [Eq.~\eqref{gapDnorm}] between Gaussian discord and optimal discord (dashed curve) and non-Gaussianity $\delta_0$ (solid curve) of the  CV Werner state \eqref{rho0}, plotted as functions of (a) the probability $p$ for different values of $\lambda$ ($\lambda = 0.2$ and $0.8$ from bottom to top), and of (b) the squeezing factor $\lambda$ (at $p=0.5$). The plots are in logarithmic scale.  All the quantities plotted are dimensionless.}
\label{figap}
\end{figure*}

To shine light on the reasons behind this, we compare the gap \begin{equation}\label{gapD}
\Delta{\cal D} \equiv {\cal D}^{G}(\rho_{0})-{\cal D}(\rho_{0})\,,\end{equation} quantifying how much Gaussian measurements overestimate the actual amount of nonclassical correlations, with a measure of the non-Gaussianity $\delta_0$ of the state \eqref{rho0} \cite{Genoni_08}. The latter is defined as the quantum relative entropy between $\rho_0$ and a reference Gaussian state $\tau_0$ with the same first and second moments,
\begin{equation}\label{nongauss}
\delta_0=\mathcal{S}\left(\tau_0\right)-\mathcal{S}\left(\rho_0\right)
\end{equation}
where $\mathcal{S}\left(\rho_0\right)$ is given in \eqref{srho0}. To find $\mathcal{S}\left(\tau_0\right)$, we note that $\rho_0$ has zero first moments but possesses a covariance matrix of the form
\begin{equation}\label{gamma0}
\Gamma_0=\left(\begin{array}{cccc}C&0&S&0\\0&C&0&-S\\S&0&C&0\\0&-S&0&C\\\end{array}\right)
\end{equation}
with $C=p\cosh(2r)+(1-p)$, $S=p\sinh(2r)$.
The von Neumann entropy of the Gaussian state $\tau_0$ can be then written as \cite{Holevo_01}
\begin{equation}\label{enttau0}
{\cal S}(\tau_0) = (\nu+1) \ln\left(\frac{\nu+1}{2}\right) - (\nu-1) \ln\left(\frac{\nu-1}{2}\right)\,,
\end{equation}
where $\nu = \sqrt{[{1-(1-2 p)^2 \lambda ^2}]/({1-\lambda ^2})}$ is the doubly-degenerate symplectic eigenvalue \cite{extremal} of the covariance matrix $\Gamma_0$.
The non-Gaussianity $\delta_0$ is a concave function of $p$ and increases with $\lambda$, diverging in the limit of infinite squeezing. In the regime of low squeezing, $\lambda \ll 1$, a series expansion (up to the quadratic term in $\lambda$) returns an approximate expression for the non-Gaussianity,
$$\delta_0^{(\lambda \ll 1)} \approx (-1+p) p \lambda ^2 [-1+\ln(p(1-p))+2\ln \lambda].$$

On the other hand, the gap in discord just coincides with the conditional entropy \eqref{Halpha} corresponding to homodyne detection, $\Delta{\cal D} = {\cal H}_{\{\Pi(\alpha)\}}^{G}(A|B)$. In the low squeezing regime, we can also expand in series (up to the quadratic term in $\lambda$) the integrand in \eqref{Halpha}, so as to obtain an approximate analytic expression for the gap,
$$\Delta{\cal D}^{(\lambda \ll 1)} \approx \pi^{-1}(-1+p) p \lambda^2 [1-\gamma-\ln2+\ln(p(1-p))+2 \ln \lambda],$$
where $\gamma \approx 0.577$ is Euler's constant.
Defining the ratio
\begin{equation}\label{fattore}
\Phi_\lambda=\frac{\pi  \left[\ln \left(\frac{4}{\lambda ^2}\right)+1\right]}
{\ln    \left(\frac{8}{\lambda ^2}\right)+\gamma -1}
\end{equation} between the approximate expressions for $\delta_0$ and $\Delta{\cal D}$ (at $p=0.5$), we see that the linearly dependent relationship
$\delta_0 \approx \Phi_\lambda \Delta{\cal D}$

holds with good approximation for small $\lambda$.
In other words, for low squeezing (say $\lambda \lesssim 0.2$), the (normalized) gap
\begin{equation}\label{gapDnorm}
\Delta{\cal D}_{\text{norm}} = \Phi_\lambda \Delta{\cal D}
\end{equation}
between optimal Gaussian (homodyne) and globally optimal non-Gaussian (photon counting) measurements for the extraction of nonclassical correlations, correctly characterizes and quantitatively reproduces the non-Gaussianity $\delta_0$ of the considered state \eqref{rho0}. Interestingly,
$$
\lim_{\lambda \rightarrow 0} \frac{\delta_0}{\Delta{\cal D}} \equiv \Phi_0 = \pi\,.
$$
This intriguing connection between the nonclassicality gap and non-Gaussianity fails to hold for larger values of $\lambda$; the discrepancy between the two parameters becomes extreme in the limit $\lambda \rightarrow 1$, when the non-Gaussianity diverges while the gap $\Delta{\cal D}$ closes to zero. A comprehensive comparison between the normalized gap in discord \eqref{gapDnorm}
and the non-Gaussianity measure \eqref{nongauss} is shown in Fig.~\ref{figap}.

\subsection{Finite versus infinite-dimensional POVMs}

One may argue, that even a simpler non-Gaussian state than
that of given in Eq.~(\ref{rho0}) can be found possessing a
strictly lower discord for a non-Gaussian measurement than for the
best Gaussian measurement. For instance, the optimal measurement minimizing the discord in the qubit Werner state \cite{werner}, studied in the seminal paper on quantum discord \cite{Olivier_01}, is a simple non-Gaussian projection onto the first two Fock states $|0\rangle$ and $|1\rangle$. One can easily check that the optimization over all Gaussian measurements gives a strictly higher discord. Let us stress that the nonclassical correlations
captured by discord are fundamentally different for the CV Werner state (\ref{rho0}) considered here and for the
qubit Werner state. Namely, although our
CV Werner \cite{Mista_02} is formally a qubit-like state, the globally optimal POVM
has an infinite number of elements given by projectors onto all Fock
states. Moreover, it is not difficult to show that no POVM
measurement on mode $B$ possessing a finite number $N$ of elements
$\Omega_{i}$, $i=1,\ldots,N$ can nullify the conditional entropy
${\cal H}(A|B)$ and hence also be globally optimal. Namely, the
conditional state corresponding to detection of the element
$\Omega_{i}$ has to be a pure state, that is,
$\rho_{A|i}^{\Omega}=\mbox{Tr}_{B}\left[\rho_{0}\Omega_{i}\right]=|\chi_{i}\rangle\langle\chi_{i}|$
in order for the entropy of the conditional state to vanish. Now,
consider the element
$\Omega_{0}\equiv\openone_{B}-\sum_{i=1}^{N}\Omega_{i}$. The
corresponding conditional state
$\rho_{A|0}^{\Omega}=\mbox{Tr}_{B}\left[\rho_{0}\Omega_{0}\right]=\rho_{0,A}-\sum_{i=1}^{N}|\chi_{i}\rangle\langle\chi_{i}|$,
where $\rho_{0,A}$ is obtained from Eq.~(\ref{rho0B}) by replacing $B$ with $A$,
cannot have neither zero, nor one, nor even any finite
number of strictly positive eigenvalues, as this would imply that the state $\rho_{0,A}$ also has a finite number of strictly
positive eigenvalues which is not the case [see equation \eqref{nurho0B}].
 Therefore, for any finite $N$ the conditional
state $\rho_{A|0}^{\Omega}$ is definitely a mixed state possessing
strictly positive entropy and hence resulting in a strictly positive and
therefore suboptimal conditional entropy (i.e., non-optimal discord).
Consequently, the globally optimal POVM of the qubit Werner state is only two-component and
thus the qubit Werner state is only a trivial embedding of a
two-qubit state into an infinitely-dimensional two-mode state space carrying {\it
only} qubit-type non-Gaussian nonclassical correlations. In contrast,
the CV Werner state (\ref{rho0}) carries genuinely CV
non-Gaussian nonclassical correlations that can be optimally
extracted only by a {\it non-Gaussian} POVM measurement with an {\it
infinite} number of elements: in this particular case, photon counting.

\subsection{Extension to mixtures of $n$ Gaussian states}

Before going further let us note that, for a state of the form (\ref{sigma}), the
global optimality of photon counting (for the calculation of quantum discord) follows immediately from the
fact that the projection of one of its modes onto a Fock state projects
the other mode onto the same Fock state. As a consequence the
conditional entropy achieves the minimum possible value
${\cal H}_{\{\Pi\left(m\right)\}}(A|B)=0$ and the discord is of the form (\ref{Drho0}).
It is not difficult to find more general mixed states with the same
property, for instance, states with the structure
\begin{equation}\label{rhoq}
\rho_{q}=\sum_{m,n=0}^{\infty}q_{mn}|mm\rangle\langle nn|,
\end{equation}
with $q_{nm}^{\ast}=q_{mn}$, $\sum_{m=0}^{\infty}q_{mm}=1$ and the matrix $Q$ with
elements $q_{ij,kl}=q_{ik}\delta_{ij}\delta_{kl}$ being positive-semidefinite.

A particular example of such states is a convex mixture of an arbitrary number of two-mode squeezed vacua (\ref{psinopa})
with different squeezing parameters $\lambda_{i}$ obtained for
$q_{mn}=\sum_{i}p_{i}\left(1-\lambda_{i}^{2}\right)\lambda_{i}^{m+n}$, where $p_{i}$ are
probabilities and $\lambda_{i}\ne\lambda_{j}$ for $i\ne j$.
We remark that for {\it all} non-Gaussian states of the form \eqref{rhoq} the quantum discord can be computed exactly.
\subsection{Other nonclassicality indicators}

The quantum state (\ref{rho0}) admits furthermore an analytical calculation of other optimized entropic quantifiers
of nonclassical correlations \cite{cavesrossignoli}, encompassing the `ameliorated' measurement-induced disturbance \cite{Girolami_10}
and the relative entropy of quantumness \cite{req,Piani_11}. Interestingly, as we will now show, both of them coincide
with the discord (\ref{Drho0}). This is in complete analogy with the corresponding Werner state for two qubits \cite{Luo_08a,discordqubits,Girolami_10} and for two qu$d$its with arbitrary $d$ \cite{wernerqudits}.

\subsubsection{Measurement-induced disturbance}

The optimized nonclassicality indicator called `ameliorated' measurement-induced disturbance (AMID) is a two-way measure of quantum correlations defined as \cite{Wu_09,Girolami_10,gamid}
\begin{equation}\label{AMID}
{\cal A}\left(\rho_{0}\right)={\cal I}_{q}\left(\rho_{0}\right)-{\cal I}_{c}(\rho_{0}),
\end{equation}
where ${\cal I}_{q}\left(\rho_{0}\right)={\cal S}(\rho_{0,A})+{\cal S}(\rho_{0,B})-{\cal S}(\rho_{0})$
is the quantum mutual information and
\begin{eqnarray}\label{Ic}
{\cal I}_{c}(\rho_{0})=\sup_{\Pi_{A}\otimes\Pi_{B}}{\cal I}(p_{AB})
\end{eqnarray}
is the classical mutual information \cite{classmut} of the quantum state $\rho_{0}$.
Here ${\cal I}(p_{AB})={\cal H}(p_{A})+{\cal H}(p_{B})-{\cal H}(p_{AB})$ is the classical mutual information of the joint probability
distribution $p_{AB}(k,l)=\mbox{Tr}[\rho_{0}\Pi_{A}(k)\otimes\Pi_{B}(l)]$ of outcomes of local
measurements $\Pi_{A}$ and $\Pi_{B}$ on $\rho_{0}$, where ${\cal H}(p_{AB})$ and ${\cal H}(p_{A})$ $({\cal H}(p_{B}))$ are
the Shannon entropies of the joint distribution and marginal distribution $p_{A}(k)=\sum_{l}p_{AB}(k,l)$ ($p_{B}(l)=\sum_{k}p_{AB}(k,l)$), respectively. The classical mutual information is upper bounded as \cite{Wu_09}
\begin{eqnarray}\label{IABbound}
{\cal I}(p_{AB})\leq\mbox{min}\left\{{\cal S}(\rho_{0,A}),{\cal S}(\rho_{0,B}),{\cal I}_{q}\left(\rho_{0}\right)\right\}.
\end{eqnarray}
Since ${\cal S}(\rho_{0,A})={\cal S}(\rho_{0,B})$ for the state \eqref{rho0}, we need to
compare ${\cal S}(\rho_{0,B})$ with ${\cal
I}_{q}\left(\rho_{0}\right)$ which can be done using majorization theory for
infinite-dimensional density matrices \cite{Wehrl_74}. Consider two such density
matrices ${\bm A}$ and ${\bm B}$ with $a_{1},a_{2},\ldots$ and $b_{1},b_{2},\ldots$
being their non-zero eigenvalues arranged in a decreasing order
and repeated according to their multiplicity. If some of the matrices, for example, ${\bm A}$
has only a finite number $k$ of non-zero eigenvalues, we set $a_{k+1}=a_{k+2}=\ldots=0$.
We then say that ${\bm A}$ is more mixed than ${\bm B}$, and write ${\bm A}\succ{\bm
B}$, if
\begin{equation}\label{majorization}
\sum_{i=1}^{k}a_{i}\leq\sum_{i=1}^{k}b_{i},\quad k=1,2,\ldots.
\end{equation}
It holds further \cite{Wehrl_74}, that if ${\bm A}\succ{\bm B}$
then their von Neumann entropies satisfy ${\cal S}({\bm A})\geq{\cal S}({\bm B})$.
Taking now the eigenvalues $\tilde{\nu}_{j}$, $j=0,1,\ldots$ given in
Eq.~(\ref{nurho0B}) instead of eigenvalues $a_{i}$, and $\nu_{1,2}$
given in Eq.~(\ref{nurho0}) instead of eigenvalues $b_{i}$, one easily
finds that they satisfy Eq.~(\ref{majorization}). This implies
that the density matrices $\rho_{0}$ and $\rho_{0,B}$ of
Eqs.~(\ref{rho0}) and (\ref{rho0B}) satisfy
$\rho_{0,B}\succ\rho_{0}$ and therefore ${\cal
S}(\rho_{0,B})\geq{\cal S}(\rho_{0})$. Hence, one gets ${\cal
I}_{q}\left(\rho_{0}\right)\geq{\cal S}(\rho_{0,B})$ which leads,
using the inequality (\ref{IABbound}), to the upper bound on the
classical mutual information in the form ${\cal I}(p_{AB})\leq{\cal
S}(\rho_{0,B})$. Considering now photon counting on both modes in the state (\ref{rho0}),
one gets ${\cal H}(p_{A})={\cal H}(p_{B})={\cal H}(p_{AB})={\cal S}(\rho_{0,A})={\cal S}(\rho_{0,B})$
which gives the classical mutual information in the form ${\cal I}(p_{AB})={\cal S}(\rho_{0,B})$.
Hence, the latter inequality is saturated by photon counting
which finally yields ${\cal A}\left(\rho_{0}\right)={\cal
S}(\rho_{0,B})-{\cal S}(\rho_{0})=\mathcal{D}(\rho_{0})$, that is, AMID coincides
with the quantum discord, \eq{Drho0}.

\subsubsection{Relative entropy of quantumness}

The measure quantifies the minimum distance, in terms of relative entropy, between a quantum state $\rho_0$ and the set of completely classically correlated states \cite{req}. It is operationally associated to the amount of distillable entanglement that can be generated between a quantum state $\rho_{0}$ (carrying nonclassical correlations) and a set of ancillary systems in the worst
case scenario of an activation protocol \cite{Piani_11}. The relative entropy of quantumness is defined as \cite{req,Piani_11}
\begin{equation}\label{Q}
{\cal Q}\left(\rho_{0}\right)=\min_{\{\Pi_{A}\otimes\Pi_{B}\}}\left[{\cal H}\left(p_{AB}\right)-{\cal S}\left(\rho_{0}\right)\right],
\end{equation}
where the minimization is performed over local measurements $\Pi_{A}$ and $\Pi_{B}$ consisting of collections of one-dimensional
projectors $\{\Pi_{A}(k)\}$ and $\{\Pi_{B}(l)\}$.

The minimum in Eq.~(\ref{Q}) can be found using the same trick as in the previous cases. First, we take a
suitable tight bound on the quantity that is to be optimized and then we guess a measurement saturating the bound.
In the present case we look for the minimum of the joint Shannon entropy ${\cal H}\left(p_{AB}\right)$ satisfying the
inequalities ${\cal H}\left(p_{AB}\right)\geq{\cal H}\left(p_{B}\right)\geq{\cal S}\left(\rho_{0,B}\right)$. Using the
previous result that for photon counting we get ${\cal H}(p_{AB})={\cal S}(\rho_{0,B})$, we see that the measurement saturates the
lower bound and therefore ${\cal Q}\left(\rho_{0}\right)={\cal S}(\rho_{0,B})-{\cal S}\left(\rho_{0}\right)$.
Thus all three considered nonclassicality indicators coincide for the state (\ref{rho0}), i.e.,
\begin{equation}\label{equality}
\mathcal{D}(\rho_{0})={\cal A}\left(\rho_{0}\right)={\cal Q}\left(\rho_{0}\right)={\cal
S}(\rho_{0,B})-{\cal S}(\rho_{0}).
\end{equation}
\section{General case}\label{secgeneral}

Let us now move to the analysis of nonclassical correlations in a
generic CV Werner state (\ref{rho}) with $\mu\ne0$,
complementing the seminal
analysis of nonclassical correlations in a two-qubit Werner state
performed in \cite{Olivier_01,Luo_08a,discordqubits,Girolami_10}.
This can be interesting in particular because, in contrast to the
qubit case, there can potentially exist PPT entangled CV Werner states \cite{Mista_02}.

In the present general case we do not have any tight bounds, similar to those of
the previous Section, allowing us to perform exact optimizations in Eqs.~(\ref{D2}),
(\ref{Ic}) and (\ref{Q}). For this reason, we cannot prove the global optimality of photon counting or any other measurement strategy analytically. We then resort to computing upper bounds on discord, AMID\footnote{In this case the upper bound on AMID is simply the nonoptimized measurement-induced disturbance (MID) \cite{Luo_08a}.} and relative entropy of quantumness, obtained for (possibly nonoptimized) measurements in the local eigenbasis of the reduced state(s) of the two-mode CV Werner state. Interestingly, all the upper bounds on the different quantities again coincide as we show later in this Section: this hints at the conjecture that they might be indeed tight for the considered states, although we cannot provide conclusive evidence of this claim. We also derive nontrivial lower bounds for the nonclassical correlations.

\subsection{Upper and lower bounds on discord}

We consider a nonoptimized upper bound on quantum discord defined for
a density matrix $\rho$ as
\begin{equation}\label{U}
\mathcal{U}(\rho)={\cal S}(\rho_{B})-{\cal
S}(\rho)+{\cal
H}_{\rm eig}(A|B),
\end{equation}
where ${\cal H}_{\rm eig}(A|B)$ is the conditional entropy
for the measurement of mode $B$ in the local eigenbasis
of the reduced state $\rho_{B}$ (see also \cite{Luo_08a}). For the general CV Werner state of \eqref{rho}, tracing
$\rho$ over mode $A$ gives the reduced state
\begin{equation}\label{rhoB}
\rho_{B}=p\rho_{B}^{\rm th}\left(\lambda\right)+(1-p)\rho_{B}^{\rm th}\left(\mu\right),
\end{equation}
where $\rho^{\rm th}$ is defined in Eq.~(\ref{rhoth}). This state
is diagonal in the Fock basis with eigenvalues
\begin{equation}\label{gm}
g_m=p\left(1-\lambda^2\right)\lambda^{2m}+\left(1-p\right)\left(1-\mu^2\right)\mu^{2m}
\end{equation}
that give, after substitution into Eq.~(\ref{S}), the marginal entropy
${\cal S}(\rho_{B})$ appearing in Eq.~(\ref{U}).

The local eigenbasis is a Fock basis and so the projection on it corresponds again, even in the present general case, to photon counting.
The conditional state
$\rho_{A|m}=\mbox{Tr}_{B}\left[|m\rangle_{B}\langle
m|\rho\right]/p_{B}(m)$, where $p_{B}(m)=\!_B\langle m|\rho_{B}|m\rangle_{B}$,
obtained by projecting mode $B$ onto Fock state $|m\rangle$ reads
explicitly
\begin{equation}\label{rhoAm}
\rho_{A|m}=\frac{p\left(1-\lambda^2\right)\lambda^{2m}|m\rangle_{A}\langle
m|+(1-p)(1-\mu^2)\mu^{2m}\rho_{A}^{\rm th}\left(\mu\right)}{p_{B}(m)}
\end{equation}
with
$p_{B}(m)=p\left(1-\lambda^2\right)\lambda^{2m}+(1-p)(1-\mu^2)\mu^{2m}$.
It has the eigenvalues
\begin{eqnarray}\label{etamn}
\eta_{n}^{(m)}&=&\frac{p\left(1-\lambda^2\right)\lambda^{2m}\delta_{mn}+(1-p)(1-\mu^2)^{2}\mu^{2(m+n)}}{p_{B}(m)},\nonumber\\
\end{eqnarray}
where $\delta_{mn}$ is the Kronecker symbol, that give the following entropy of the conditional state
\begin{eqnarray}\label{SrhoAm}
\mathcal{S}\left(\rho_{A|m}\right)&=&-\frac{\left(1-p\right)\left(1-\mu^2\right)^{2}\mu^{2m}}{p_{B}\left(m\right)}\nonumber\\
&&\times\left\{\ln\left[\frac{\left(1-p\right)
\left(1-\mu^2\right)^{2}}{p_{B}(m)}\right]\right.\nonumber\\
&&\left.\times\left(\frac{1}{1-\mu^{2}}-\mu^{2m}\right)+\ln\left(\mu^2\right)\right.\nonumber\\
&&\left.\times\left[\frac{m}{1-\mu^2}+\frac{\mu^2}{\left(1-\mu^2\right)^{2}}-2m\mu^{2m}\right]\right\}\nonumber\\
&&-\eta_{m}^{(m)}\ln\eta_{m}^{(m)}.
\end{eqnarray}
Hence, one gets the conditional entropy ${\cal H}_{\rm eig}(A|B)=\sum_{m=0}^{\infty}p_{B}(m)\mathcal{S}\left(\rho_{A|m}\right)$.

It remains to calculate the global entropy of the state (\ref{rho}).
For this purpose it is convenient to express the state as
\begin{equation}\label{rhogen}
\rho=\sum_{m,n=0}^{\infty}M_{mn}|m,m\rangle\langle n,n|+\sum_{m\ne
n=0}^{\infty}e_{mn}|m,n\rangle\langle m,n|,
\end{equation}
where
\begin{eqnarray}
M_{mn}&=&p\left(1-\lambda^2\right)\lambda^{m+n}\nonumber\\
&&+\left(1-p\right)\left(1-\mu^2\right)^2\mu^{2(m+n)}\delta_{mn},\label{Mmn}\\
e_{mn}&=&\left(1-p\right)\left(1-\mu^2\right)^2\mu^{2(m+n)}.
\end{eqnarray}
The state (\ref{rho}) thus possesses the eigenvalues $e_{mn}$ corresponding to the
eigenvectors $|m,n\rangle$, $m\ne n=0,1,\ldots$ and the remaining
eigenvalues $\left(\equiv f_{l}\right)$ are the eigenvalues of the
infinite-dimensional matrix $M$ with elements (\ref{Mmn}). This gives the global entropy
\begin{eqnarray}\label{globalS}
\mathcal{S}\left(\rho\right)&=&-\frac{2\mu^2\left(1-p\right)}{1+\mu^2}\left\{\ln\left[\left(1-p\right)\left(1-\mu^2\right)^2\right]\right.
\nonumber\\
&&\left.+\frac{2\ln\left(\mu\right)\left(1+\mu^2+2\mu^4\right)}{1-\mu^4}\right\}
-\sum^\infty_{l=0}f_l\ln f_{l}.\nonumber\\
\end{eqnarray}
The eigenvalues $f_{l}$ of matrix $M$ appearing in the last expression of the previous equation
cannot be calculated analytically and one has to resort to numerical diagonalization of a
sufficiently large truncated matrix. Hence, one gets using Eq.~(\ref{globalS}), and expressions
for ${\cal H}_{\rm eig}(A|B)$ and ${\cal S}(\rho_{B})$, the sought upper bound (\ref{U}) on the true quantum discord.

The true discord can be also bounded from below in the following way. Let us observe first, that
apart from the trivial case $p=0$ (corresponding to a product state) all other CV Werner states have nonclassical correlations as they possess a
strictly positive quantum discord ${\cal D}(\rho)>0$. This can be proven using the sufficient condition on
strict positivity of quantum discord \cite{Bylicka_11} according to which ${\cal D}(\rho)>0$ for a state
$\rho$ if at least one off-diagonal block $\rho_{ij}^{(B)}\equiv\!_{B}\langle i|\rho|j\rangle_{B}$, $i\ne j$
is not normal, i.e., it does not commute with its adjoint. In the present case of the Werner state (\ref{rho}) we
have explicitly $\rho_{ij}^{(B)}=p\left(1-\lambda^2\right)\lambda^{i+j}|i\rangle_{B}\langle j|$.
Assuming $p>0$ and $0<\lambda<1$ this gives immediately a nonzero commutator
\begin{eqnarray}\label{commutator}
\left[\rho_{ij}^{(B)},\left(\rho_{ij}^{(B)}\right)^{\dag}\right]&=&p^2\left(1-\lambda^2\right)^2\lambda^{2(i+j)}
\left[|i\rangle_{B}\langle i|-|j\rangle_{B}\langle j|\right]\nonumber\\
&\ne& 0 \quad \mbox{for}\,\,\, i\ne j, \, \, p\neq0,
\end{eqnarray}
as required.

An explicit, non-tight lower bound can be derived that is nonnegative (and thus nontrivial) at least on some subinterval of probabilities $p$. Namely, assume a POVM on mode $B$ given by a collection of rank-1 operators $\left\{|\psi_{j}\rangle\langle\psi_{j}|\right\}$.
If the component $|\psi_{i}\rangle\langle\psi_{i}|$ is detected on mode $B$ in the state (\ref{rho}), then
mode $A$ collapses into the normalized conditional state
\begin{equation}\label{psicond}
\rho_{A|i}=\frac{p|\phi_{i}\rangle_{A}\langle
\phi_{i}|+(1-p)\langle\psi_{i}|\rho_{B}^{\rm th}\left(\mu\right)|\psi_{i}\rangle\rho_{A}^{\rm th}\left(\mu\right)}{p_i},
\end{equation}
where $|\phi_{i}\rangle_{A}$ is a pure unnormalized state that is not specified here and
$p_{i}=\langle\psi_{i}|\rho_{B}|\psi_{i}\rangle$ with $\rho_{B}$ given in Eq.~(\ref{rhoB}) is
the probability of measuring the outcome $i$.
The state is a convex mixture of a pure state and a thermal state. Making use of the concavity of the von Neumann entropy
${\cal S}\left(\sum_{j}p_{j}\rho_{j}\right)\geq\sum_{j}p_{j}{\cal S}\left(\rho_{j}\right)$
and the fact that the entropy vanishes on pure states we arrive at the following inequality:
\begin{equation}\label{inequality}
\mathcal{S}\left(\rho_{A|i}\right)\geq\frac{(1-p)}{p_{i}}\langle\psi_{i}|\rho_{B}^{\rm th}\left(\mu\right)|\psi_{i}\rangle\mathcal{S}\left[\rho_{A}^{\rm th}\left(\mu\right)\right].
\end{equation}
By multiplying both sides of the inequality by $p_{i}$ and summing over $i$ one finds
the classical conditional entropy to be lower bounded as
${\cal H}_{\{|\psi_{i}\rangle\langle\psi_{i}|\}}(A|B)\geq(1-p)\mathcal{S}\left[\rho_{A}^{\rm th}\left(\mu\right)\right]$
which yields finally using Eq.~(\ref{D2}) the lower bound
\begin{equation}\label{L}
\mathcal{L}(\rho)={\cal S}\left(\rho_{B}\right)-{\cal S}\left(\rho\right)+\left(1-p\right){\cal S}\left[\rho_{A}^{\rm th}\left(\mu\right)\right].
\end{equation}

In what follows we evaluate the upper and lower bounds in Eqs.~(\ref{U}) and (\ref{L}), respectively,
for two particularly important two-parametric subfamilies of the set of CV Werner states.
\subsection{Examples}

\subsubsection{The case $\lambda=\mu$}

\begin{figure}[tb]
\includegraphics[width=7.5cm]{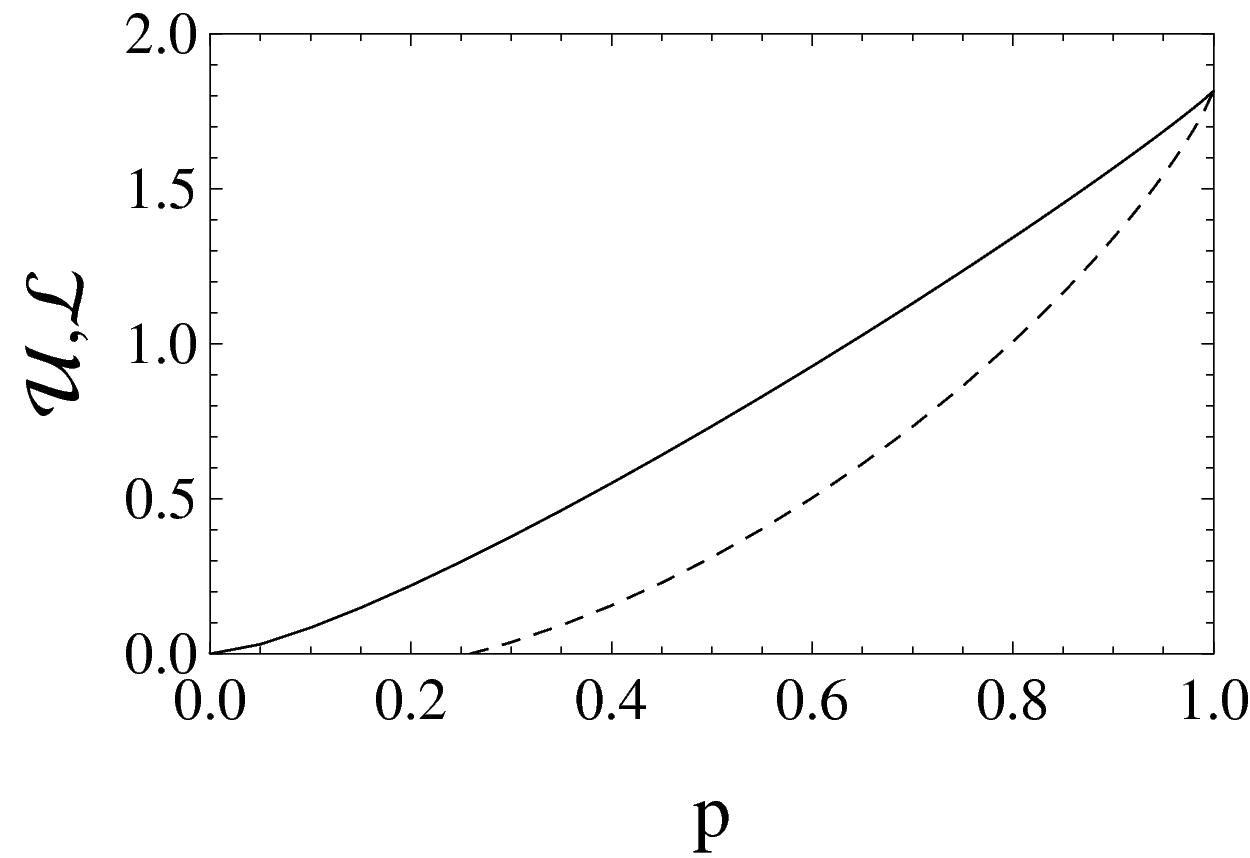}
\caption{Upper bound $\mathcal{U}$, Eq.~(\ref{U}) (solid curve), and lower bound $\mathcal{L}$, Eq.~(\ref{L}) (dashed curve),
on quantum discord (\ref{D2}) versus probability $p$ for the CV Werner state (\ref{rho}) with $\lambda=\mu=0.8$. All the quantities plotted are dimensionless.} \label{fig2}
\end{figure}

First, we consider the case $\lambda=\mu$. Then
the reduced state (\ref{rhoB}) is just a thermal state
$\rho_{B}=\rho_{B}^{\rm th}(\lambda)$ with a well-known entropy
\begin{equation}\label{rhoAent}
{\cal
S}(\rho_{B})=-\frac{\ln\left(1-\lambda^{2}\right)}{\left(1-\lambda^{2}\right)}-\frac{\lambda^{2}}{1-\lambda^{2}}\ln
\left(\frac{\lambda^2}{1-\lambda^{2}}\right).
\end{equation}
In this case the CV Werner state is a mixture of a pure two-mode squeezed vacuum state (\ref{psinopa}) with a product of its marginals; in the strong squeezing limit $\lambda \rightarrow 1$, this state approaches a mixture of a maximally entangled EPR
state and a maximally mixed (infinitely thermal) state, which is a direct CV counterpart to the qubit Werner state \cite{werner}.
Further, it can be shown that the CV Werner state (\ref{rho}) with $\lambda=\mu>0$ is entangled for
any $p>0$ \cite{Mista_02}. The upper bound (\ref{U}) and lower bound (\ref{L}) on quantum discord are depicted
in Fig.~\ref{fig2}. Note that in this and in the following plots, only nonzero values of the lower bound (\ref{L}) will be shown.

\subsubsection{The case $\lambda=\mu^4$}

\begin{figure}[tb]
\includegraphics[width=7.5cm]{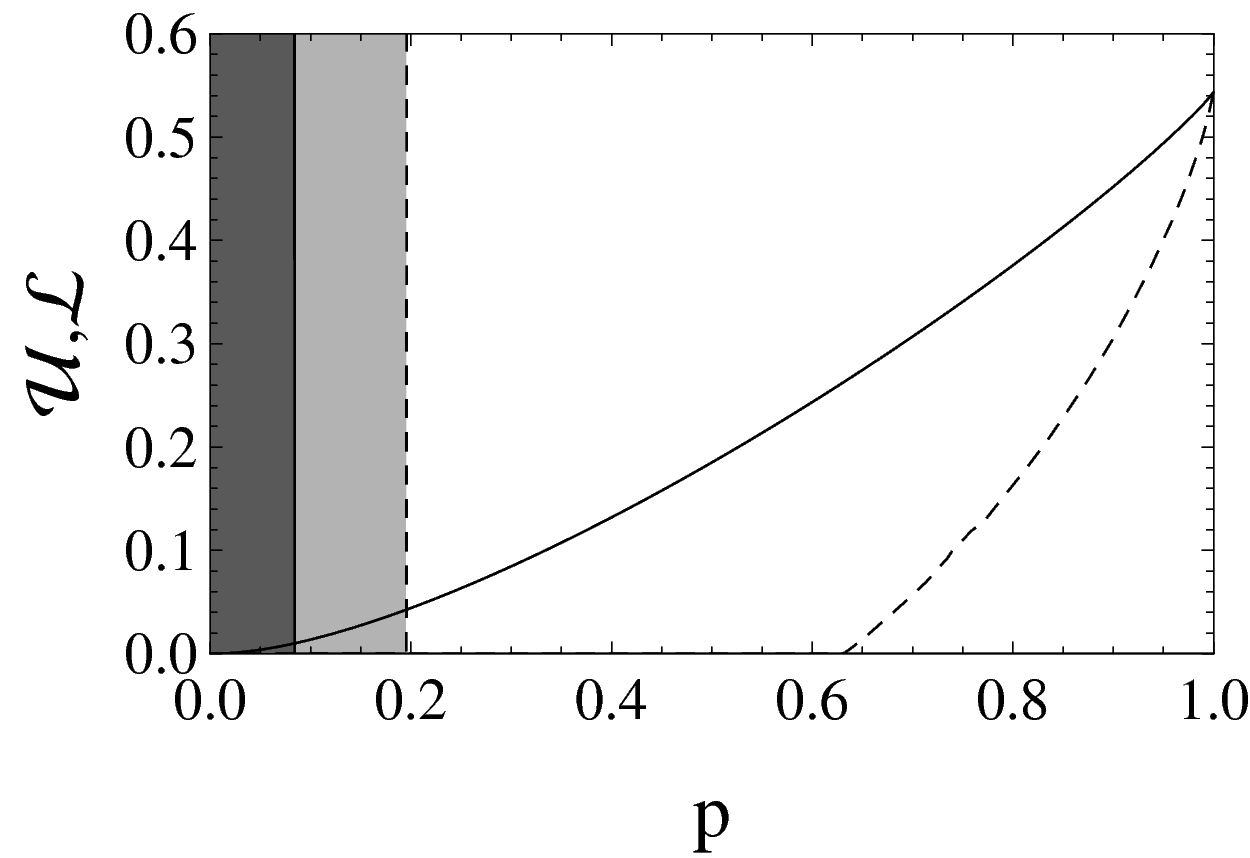}
\caption{Upper bound $\mathcal{U}$, Eq.~(\ref{U}) (solid curve), and lower bound $\mathcal{L}$, Eq.~(\ref{L}) (dashed curve),
on quantum discord (\ref{D2}) versus probability $p$ for the CV Werner state (\ref{rho}) with $\lambda=\mu^4$ and $\mu=0.8$.
The dark gray shaded region corresponds to separable states, the light gray shaded region corresponds to PPT states with unknown
separability properties, and the white region corresponds to entangled non-PPT states. The boundary probabilities $p_{\rm sep}$ and $p_{\rm PPT}$ are depicted by vertical solid line and dashed line, respectively. All the quantities plotted are dimensionless.} \label{fig3}
\end{figure}

The case for which $\lambda=\mu^4$ is interesting because the CV
Werner state (\ref{rho}) runs through three different
separability regions as the parameter $p$ increases \cite{Mista_02}:
\begin{enumerate}
\item If $p\leq p_{\rm
sep}\equiv\frac{(1-\mu^{2})^{2}}{2(1-\mu^{2}+\mu^{4})}$, then the
state $\rho$ is separable (dark gray strip in Fig.~\ref{fig3}). \item
If $p_{\rm sep}<p\leq p_{\rm
PPT}\equiv\frac{\left(1-\mu^2\right)^2}{\left(1-\mu^2\right)^2+\left(1-\mu^8\right)\mu^2}$,
 the state $\rho$ is PPT (i.e., it has positive partial transposition \cite{PPT}) and it is unknown whether it is undistillable-entangled or separable (light gray strip in
Fig.~\ref{fig3}). \item If $p>p_{\rm PPT}$, then the state $\rho$
is non-PPT and therefore entangled (white region in Fig.~\ref{fig3}).
\end{enumerate}

\subsection{Other nonoptimized nonclassicality indicators}


Next we focus on the determination of the nonoptimized (upper bound) version of AMID (\ref{AMID}) called
measurement-induced disturbance (MID) defined as \cite{Luo_08a}
\begin{equation}\label{MID}
\mathcal{M}(\rho)={\cal I}_{q}(\rho)-{\cal
I}(p_{AB}),
\end{equation}
where ${\cal I}(p_{AB})$ is the classical Shannon mutual information of
a probability distribution of results of measurements in eigenbases of the reduced states
$\rho_{A,B}$, which in the present case coincides with the joint photon-number distribution
$p_{AB}(m,n)=\!_{AB}\langle m n|\rho|m n\rangle_{AB}$. Its marginal distributions $p_{A}=p_{B}$ coincide
with the eigenvalues of the reduced states (\ref{gm}), i.e., $p_{A}(m)=p_{B}(m)=g_{m}$ whence we get the equality
between local Shannon and von Neumann entropies
\begin{equation}\label{SHequality}
{\cal H}\left( p_{A}\right)={\cal H}\left(p_{B}\right)={\cal S}\left(\rho_{A}\right)={\cal S}\left(\rho_{B}\right).
\end{equation}
Hence MID simplifies to
\begin{equation}\label{MIDsimple}
\mathcal{M}(\rho)={\cal H}\left(p_{AB}\right)-{\cal S}\left(\rho\right).
\end{equation}
The global Shannon entropy can be derived easily by noting that the eigenvalues (\ref{etamn})
of the conditional state (\ref{rhoAm}) satisfy $\eta_{n}^{(m)}=p_{AB}(m,n)/p_{B}(m)$ thus representing a conditional
probability $p_{AB}(n|m)$ of detecting $n$ photons in mode $A$ given $m$ photons have been detected in mode $B$. This
implies immediately that ${\cal H}_{\rm eig}(A|B)={\cal H}\left(p_{AB}\right)-{\cal S}\left(\rho_{B}\right)$,
where we have used Eq.~(\ref{SHequality}) and the equality $p_{A}(m)=p_{B}(m)$. Substituting from here for
${\cal H}\left(p_{AB}\right)$ into Eq.~(\ref{MIDsimple}) finally leads to the equality of the upper bound on discord
(\ref{U}) and MID (\ref{MID})
\begin{equation}\label{UMIDequality}
\mathcal{M}(\rho)=\mathcal{U}(\rho).
\end{equation}
Note that the two coincident quantities also provide an upper bound for the relative entropy of quantumness \cite{req,cavesrossignoli} of the states \eqref{rho}. Note further that the lower bound \eqref{L} on discord is also a lower bound for the other measures of nonclassical correlations such as AMID, MID, and relative entropy of quantumness, since quantum discord is in general smaller than those mentioned quantities for arbitrary bipartite quantum states \cite{Wu_09,cavesrossignoli}.

\section{Partially transposed CV Werner state}\label{secppt}

One of the main technical disadvantages  of the CV Werner state with $\mu\ne0$ is that its eigenvalues, and consequently its global von Neumann entropy, cannot be calculated analytically. Interestingly, this ceases to be the case if the state is partially
transposed \cite{Mista_02}. Then, one can find regions of parameters $p,\lambda$ and $\mu$ for which the
partial transposes are positive-semidefinite, and so represent a legitimate quantum state in their own right. Thus, one
obtains another, sometimes simpler to treat family of non-Gaussian quantum states for which we can get further with analytical tools than in the case of the original state.

Let us illustrate this on a simple example of the CV Werner state (\ref{rho}) with $\lambda=\mu^{2}$. As was shown in
Ref.~\cite{Mista_02} for
\begin{equation}\label{pcondition}
p=\frac{1-\lambda}{2},
\end{equation}
the partial transposition $\rho^{T_A} \equiv \tilde{\rho}$ of the Werner state $\rho$ with respect to mode $A$,
\begin{eqnarray}\label{tilderho}
\tilde{\rho}&=&{\cal N}\sum_{m,n=0}^{\infty}\lambda^{m+n}
\left(|n,m\rangle\langle m,n|+|m,n\rangle\langle m,n|\right),\nonumber\\
\end{eqnarray}
possesses the following nonnegative nonzero eigenvalues
\begin{eqnarray}\label{eigenvalues}
a_{m}&=&2{\cal N}\lambda^{2m},\quad m=0,1,\ldots,\\
b_{mn}&=&2{\cal N}\lambda^{m+n},\quad m>n=0,1,\ldots,
\end{eqnarray}
where ${\cal N}=\left(1-\lambda^{2}\right)\left(1-\lambda\right)/2$, and is therefore a valid two-mode density matrix corresponding to a different non-Gaussian state.
Direct substitution of the eigenvalues into Eq.~(\ref{S}) gives the analytical expression for the global
entropy of $\tilde{\rho}$ of the form
\begin{eqnarray}\label{Stilderho}
{\cal S}\left(\tilde{\rho}\right)&=&-\left[\ln\left(2{\cal N}\right)+
\frac{1+3\lambda}{1-\lambda^2}\lambda\ln\lambda\right].
\end{eqnarray}
Tracing the state (\ref{tilderho}) over mode $A$ one gets the reduced state of mode $B$
\begin{equation}\label{tilderhoB}
\tilde{\rho}_{B}={\cal N}\sum_{m=0}^{\infty}\left(\lambda^{2m}+\frac{\lambda^m}{1-\lambda}\right)|m\rangle_{B}\langle m|
\end{equation}
with entropy
\begin{eqnarray}\label{StilderhoB}
{\cal S}\left(\tilde{\rho}_{B}\right)&=&-\left[{\cal N}\sum_{m=0}^{\infty}\left(\lambda^{2m}+\frac{\lambda^m}{1-\lambda}\right)\ln\left(\lambda^{m}+\frac{1}{1-\lambda}\right)\right.\nonumber\\
&&+\left.\ln\left({\cal N}\right)+\frac{\lambda\left(1+3\lambda\right)}{2\left(1-\lambda^2\right)}\ln\lambda\right].
\end{eqnarray}
Similarly one can find a reduced state $\tilde{\rho}_{A}$ of mode $A$ which coincides with the reduced state (\ref{tilderhoB})
and yields the local entropy ${\cal S}\left(\tilde{\rho}_{A}\right)={\cal S}\left(\tilde{\rho}_{B}\right)$.

\begin{figure}[tb]
\includegraphics[width=7.5cm]{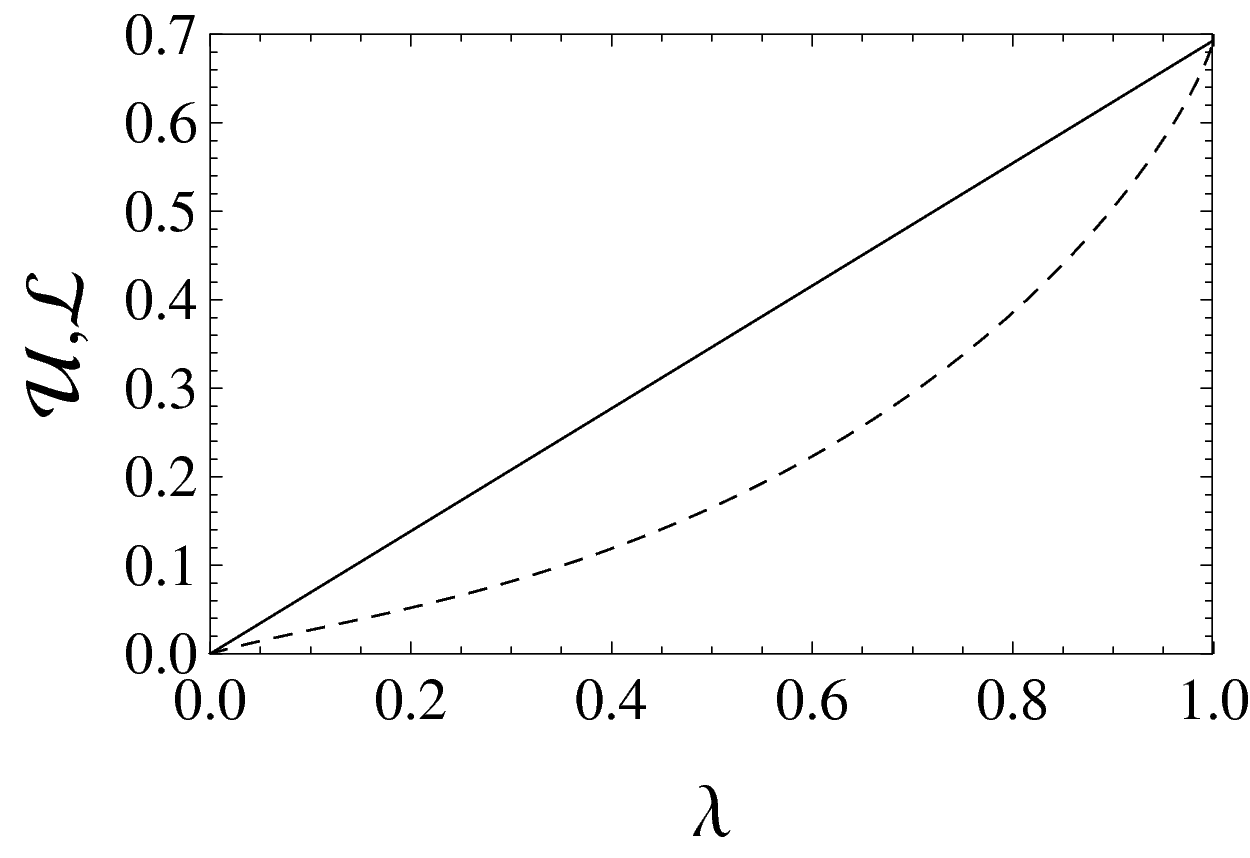}
\caption{Upper bound $\mathcal{U}$, Eq.~(\ref{Utilderho}) (solid curve), and lower bound $\mathcal{L}$, Eq.~(\ref{Ltilderho}) (dashed curve), on quantum discord (\ref{D2}) versus squeezing parameter $\lambda=\tanh r$ for the partially transposed CV Werner state (\ref{tilderho}). All the quantities plotted are dimensionless.} \label{fig4}
\end{figure}

Upon detecting $m$ photons in mode $B$ in the state (\ref{tilderho}), mode $A$ collapses
into the normalized conditional state
\begin{equation}\label{tilderhoAm}
\tilde{\rho}_{A|m}=\frac{\cal N}{\tilde{p}_{B}(m)}\left(\lambda^{2m}|m\rangle_{A}\langle
m|+\lambda^{m}\sum_{k=0}^{\infty}\lambda^{k}|k\rangle_{A}\langle
k|\right),
\end{equation}
where
\begin{equation}\label{tildepm}
\tilde{p}_{B}(m)={\cal N}\lambda^{m}\left(\lambda^{m}+\frac{1}{1-\lambda}\right)
\end{equation}
is the probability of detecting $m$ photons on mode $B$ in the state (\ref{tilderho}). After some algebra,
the corresponding nonoptimized conditional entropy
$\tilde{{\cal H}}_{\rm eig}(A|B)=\sum_{m=0}^{\infty}\tilde{p}_{B}(m)\mathcal{S}\left(\tilde{\rho}_{A|m}\right)$
attains then the form
\begin{equation}\label{tildecalH}
\tilde{{\cal H}}_{\rm eig}(A|B)={\cal S}\left(\tilde{\rho}\right)-{\cal S}\left(\tilde{\rho}_{B}\right)+\lambda\ln2,
\end{equation}
where we have used Eqs.~(\ref{Stilderho}) and (\ref{StilderhoB}). Substituting finally the latter formula into Eq.~(\ref{U})
we arrive at a very simple analytical expression for the upper bound on the quantum discord of the state \eqref{tilderho},
\begin{equation}\label{Utilderho}
{\cal U}\left(\tilde{\rho}\right)=\lambda\ln2.
\end{equation}
Even for the partially transposed CV Werner state $\tilde{\rho}$, it is possible to derive a nontrivial lower bound on
quantum discord (\ref{D2}). Repeating the algorithm leading to Eq.~(\ref{L}) for the state
(\ref{tilderho}), one gets the lower bound in the form:
\begin{equation}\label{Ltilderho}
\mathcal{L}(\tilde{\rho})={\cal S}\left(\tilde{\rho}_{B}\right)-{\cal S}\left(\tilde{\rho}\right)+\left(\frac{1+\lambda}{2}\right){\cal S}\left[\rho_{A}^{\rm th}\left(\sqrt{\lambda}\right)\right].
\end{equation}
The upper bound (\ref{Utilderho}) and lower bound (\ref{Ltilderho}) on discord  are depicted in Fig.~\ref{fig4} as a function of the parameter $\lambda$.  Note that in this case they are quite close to each other, with the lower bound being always faithful in the whole considered parameter space. Also, nonclassical correlations in this PPT state (which may be separable or at most contain undistillable entanglement) are quite weak (yet always nonzero), increasing slowly with the squeezing $r$ (recall that $\lambda=\tanh r$) and converging to the small, finite value $\ln 2$ in the limit $r \rightarrow \infty$.

As for other nonclassicality indicators, we get equivalent results. Moving for instance to the evaluation of the MID (\ref{MID}), one gets the joint photon-number distribution for the state
(\ref{tilderho}) to be
\begin{equation}\label{tildepmn}
\tilde{p}_{AB}\left(m,n\right)={\cal N}\lambda^{m+n}\left(1+\delta_{mn}\right).
\end{equation}
The Shannon entropy of the distribution reads
\begin{equation}\label{HtildepAB}
{\cal H}\left(\tilde{p}_{AB}\right)={\cal S}\left(\tilde{\rho}\right)+\lambda\ln2,
\end{equation}
where the global entropy ${\cal S}\left(\tilde{\rho}\right)$ is given in Eq.~(\ref{Stilderho}).
The marginal distributions on each mode coincide and they are given by Eq.~(\ref{tildepm}). Hence for the state $\tilde{\rho}$ also, the local Shannon and von Neumann entropies satisfy Eq.~(\ref{SHequality}).
Making use of the latter equality
in the definition of MID (\ref{MID}) we can express it as
\begin{equation}\label{MID2}
\mathcal{M}(\tilde{\rho})={\cal H}\left(\tilde{p}_{AB}\right)-{\cal S}\left(\tilde{\rho}\right)=\lambda\ln2,
\end{equation}
where in the derivation of the second equality we used Eq.~(\ref{HtildepAB}). Thus, as for the generic
CV Werner state of the previous Section, for the considered partially transposed CV Werner state the
MID coincides with the upper bound on discord associated with local photon counting,  i.e., $\mathcal{M}(\tilde{\rho})={\cal U}\left(\tilde{\rho}\right)$.

\section{Conclusions}

\label{secfinal}

In this paper we studied the nonclassicality of correlations in a class of non-Gaussian states of a two-mode CV system. We adopted primarily quantum discord \cite{Olivier_01,Vedral_01}, as well as other quantifiers such as measurement-induced disturbance and relative entropy of quantumness, and provided exact results (when possible) and in general upper and lower bounds on their quantification in the considered states. Our analysis enabled us to venture beyond Gaussian states and operations to access general features of quantum correlations in infinite-dimensional states. We focused in particular on two-mode CV Werner states \cite{Mista_02}, constructed as mixtures of Gaussian states and thus possessing a positive, yet non-Gaussian-shaped Wigner function on the quantum phase space. Such states provide important testbeds for the understanding of nonclassicality beyond entanglement and its interplay with entanglement itself. For the special case of a two-mode squeezed state mixed with a vacuum \cite{Lund_06}, we were able to calculate quantum discord exactly, proving that local photon counting constitutes the globally optimal strategy to extract nonclassical correlations, and any Gaussian measurement strategy turns out to be suboptimal for the task. The considered states constitute probably the simplest example of bipartite states possessing genuinely non-Gaussian CV nonclassical correlations beside entanglement. The relationship between nonclassicality and non-Gaussianity was further highlighted by observing that the gap between the quantum discord and its (non-optimal) counterpart, restricted to Gaussian measurements only, scales linearly with an entropic measure of non-Gaussianity \cite{Genoni_08} for the considered class of states in the regime of low squeezing. For general CV Werner states, photon counting measurements provide in general upper bounds on quantum discord and related nonclassicality measures, but we could not prove the tightness of such bounds analytically. We are tempted to conjecture that our upper bound does yield the true quantum discord for all CV Werner states, and future progress on validating or disproving this claim would be valuable. On the other hand, we derived nontrivial lower bounds on quantum discord as well for the considered states, which we do not expect to be tight.
We finally constructed a particular instance of a non-Gaussian state which is positive under partial transposition, and whose upper and lower bounds on discord are analytically computable and allow us to pin down its nonclassical correlations as being quite weak, staying finite even in the limit of infinite squeezing.
Our study evidences a trend for quantum correlations to be generally limited in the absence of (distillable) entanglement, as originally noted for Gaussian states \cite{gdiscord}. It would be interesting to provide tight upper bounds on the attainable amount of quantum discord for all separable CV states. Such a bound is known for two-mode Gaussian states and corresponds to one unit of discord, on a scale ranging to infinity \cite{gdiscord}. On finite-dimensional two-qu$d$it systems, on the other hand, it is known that separable mixed states can have as much nonclassical correlations as maximally entangled pure states, approaching the bound $\ln d$ when measured by the relative entropy of quantumness \cite{req}.

Nonclassicality and non-Gaussianity are two of the most important resources for the optimization and realistic implementation of present-day and next-generation quantum technology \cite{merali,nongauss}. This paper realizes a first step to explore the interplay between the two in physically relevant CV states. An important next target for future work would be to study the structure of nonclassical correlations in other practically useful states deviating from Gaussianity, such as photon-subtracted states \cite{nongausscommun,carles,iophot}. Comparing the performance of non-Gaussian measurements, such as photon counting, with that of Gaussian strategies, such as homodyne and heterodyne detection, for accessing nonclassical correlations, and studying how the gap between the two scales with the non-Gaussianity of the states \cite{Genoni_08}, and with other nonclassical parameters widely used in quantum optics \cite{opt}, could provide novel insight into the nature of quantumness (in its broadest sense) and its potential exploitation for CV quantum information.

\acknowledgments{
L. M., R. T., and N. K. acknowledge the EU grant under FET-Open project COMPAS
(212008).
L. M.  has been supported by projects ``Measurement and Information in Optics,'' (MSM 6198959213) and
Center of Modern Optics (LC06007) of the Czech Ministry of Education. R. T. and N. K. are grateful for the support from SUPA (Scottish  Universities Physics Alliance). G. A. was supported by a  Nottingham Early Career Research and Knowledge Transfer Award.}


\end{document}